# Contribution of Shorter-term Radiative Forcings of Aerosols and Ozone to Global Warming in the Last Two Decades

Qing-Bin Lu

Department of Physics and Astronomy and Departments of Biology and Chemistry, University of Waterloo, 200 University Avenue West, Waterloo, Ontario, Canada (Email: qblu@uwaterloo.ca)

**Abstract**: This paper reports observations of regional and global upper stratosphere temperature (UST) and surface temperature, as well as various climate drivers including greenhouse gases (GHGs), ozone, aerosols, solar variability, snow cover extent, and sea ice extent (SIE). We strikingly found warming trends of 0.77(±0.57) and 0.69(±0.22) K/decade in UST at altitudes of 35-40 km in the Arctic and Antarctic respectively and no significant trends over non-polar regions since 2002. These UST trends provide fingerprints of decreasing and no significant trends in total GHG effect in polar and non-polar regions respectively. Correspondingly, we made the first observation of surface cooling trends in both the Antarctic since 2005 and the Arctic since 2016 once the SIE started to recover. But surface warming remains at mid-latitudes, which causes the recent rise in global mean surface temperature (GMST). These temperature changing patterns are consistent with the characteristics of the cosmic-ray-driven electron reaction (CRE) mechanism of halogen-containing GHGs (halo-GHGs) with larger destruction rates at higher latitudes. Moreover, the no-parameter physics model of warming caused by halo-GHGs reproduces closely the observed GMSTs from 2000 to 2024, including the almost no warming during 2000-2012 and the significant warming by 0.2-0.3 °C during 2013-2023, of which 0.27 °C was calculated to arise from the net radiative forcing of aerosols and ozone due to improved air quality. The results also show that the physics model captures 76% of the variance in the observed GMSTs, exhibiting a warming peak in October 2023 and predicting a gradual GMST reversal thereafter. The results from this study may greatly improve our understanding of global climate change and lead to the identifying of the correct major culprit for human contribution to changing the climate.

**Abbreviations:** AA, Arctic amplification; AMSU, Advanced Microwave Sounding Unit; CFCs, chlorofluorocarbons; CR, cosmic ray; CRE, cosmic-ray-driven electron-induced reaction; ENSO, El Niño southern oscillation; ERF, effective radiative forcing; EUMETSAT, European Organisation for the Exploitation of Meteorological Satellites; GCM, general circulation model; GHG, greenhouse gas; GMST, global mean surface temperature; halo-GHG, halogen-containing greenhouse gas; LST, lower-stratospheric temperature; MSU, Microwave Sounding Unit; NOAA, National Oceanic and Atmospheric Administration; ODS, ozone-depleting substance; RF, radiative forcing; RO, Radio Occultation; ROM SAF, Radio Occultation Meteorology Satellite Application Facility; SCE, snow cover extent; SIE, sea ice extent; SH/NH, Southern/Northern Hemisphere; SSU, Stratospheric Sounding Unit; TSI, total solar irradiance; UST, upper stratosphere temperature; WMGHG, well-mixed greenhouse gas.

## 1. INTRODUCTION

It is well documented in the literature that conventional climate models (i.e., general circulation models – GCMs) not only include unresolved terms embodied in equations with adjustable parameters but also have major limitations such as the structural error and uncertainty across models and the requirement of tuning the models to match the observed temperatures[1, 2]. An alternative empirical approach to numerical model simulations was developed by Lean and Rind[3, 4], who reconstructed monthly global or regional mean surface temperature anomalies as $T_R(t) = \alpha_0 + \alpha_1 E(t - \Delta t_E) + \alpha_2 V(t - \Delta t_V) + \alpha_3 S(t - \Delta t_S) + \alpha_4 A(t - \Delta t_A)$, where E, V, S and A were zero mean, unit variance time series of ENSO, volcanic aerosols, solar irradiance, and anthropogenic forcing respectively, and the lags were $\Delta t_E = 4$ months, $\Delta t_V = 6$ months, $\Delta t_S = 1$ month, and $\Delta t_A = 10$ year, chosen to maximize the explained variance. The fitted parameters (coefficients), $\alpha_i$ (i=0-4), were obtained by multiple linear regression to the historical surface temperature record. They used this empirical model to assess climate changes in time scales of a few decades, showing that the combination of natural and anthropogenic forcings (at chosen lags) accounted for 76% of the variance in the monthly global surface temperature record of 1980-2008. Douglass and Clader[5] have also used a linear model to determine the climate sensitivity of the Earth to solar irradiance from multiple temperature anomalies data sets, and their results are consistent with earlier estimates of solar climate sensitivity from ocean temperatures on decadal scales and paleo-reconstructed temperatures on centennial scales. On the basis of these prior studies, the author[6] developed a conceptual physics model of climate change, 'the chlorofluorocarbon (CFC)-warming physics model'. In this CFC-warming physics model, the lags found in the Lean and Rind empirical model [3, 4], one of which (the lag of 10 years for anthropogenic forcing) is particularly in excellent agreement with ozone observations (see below), and an observational approach to quantify the solar climate forcing and sensitivity, similar to the Douglass–Clader approach[5], were adopted. In comparison, this CFC physics model has a major advantage: it applies the quantum physics of the Earth blackbody radiation to find the climate sensitivity to effective greenhouse gases (GHGs) in the atmospheric window at 8-13 μm and therefore includes no tunable (fitting) parameters. The CFC model can directly perform analytical calculations of variations in global mean surface temperature (GMST) caused by halogenated GHGs (halo-GHGs), ozone, and solar (aerosol/cloudiness) variability, provided that the radiative forcings are calculated from simple analytical formulations or GCMs. For a thorough comparison between the above-mentioned various clime models including GCMs and the no-parameter CFC-warming physics model, readers are referred to a recent review[7]. This CFC-warming physics model has predicted a long-term reversal in GMST, corresponding to the changing trend of atmospheric halo-GHGs[6-10].

Furthermore, according to the cosmic-ray (CR) driven electron-induced reaction (CRE) mechanism of ODSs for global ozone depletion[6, 8, 10-12], the destruction of ODSs or halo-GHGs by the CRE reaction is increasingly effective with increasing latitudes due to increasing CR fluxes affected by the geomagnetic field. As a direct consequence, the CFC-warming and CRE models have predicted that the reversal in upper-stratospheric temperature (UST) from cooling to warming due to the GHG effect should have occurred first at high latitudes (the polar regions), at which the levels of halo-GHGs (mainly CFCs) decrease first, and that it will lag by about 10 years over the tropics and mid-latitudes[7, 13]. But this reversal in UST is counteracted by the emerging recovery of the $O_3$ layer at high latitudes[11], as ozone itself is an effective GHG. Such a counteracting effect should be much smaller at the Northern Hemisphere (NH) than the Southern Hemisphere (SH) because $O_3$ loss over the Arctic is well known to be far less than over the Antarctic since the 1960s.

Also, there was a significant increase in tropospheric ozone at NH mid- and high-latitudes due to serious air pollution in the late half of the 20$^{th}$ century[1], while tropospheric ozone in the pollution-free Antarctic exhibited little change or some depletion associated with the CRE mechanism[12]. With continuously improving of air quality, it is expected to see a significant reversal (decrease) in tropospheric ozone at NH mid-latitudes in the coming decades[1]. A combination of these effects is expected to see more significant upper-stratospheric warming in the Arctic (and even NH mid-latitudes) than the Antarctic (and even SH mid-latitudes). Correspondingly, a reversal from warming to cooling on the surface should also first occur at high latitudes, which has been observed in the Antarctic[7] and was predicted to emerge in the Arctic once the Arctic amplification (AA) effect on surface temperature becomes insignificant, that is, once the NH sea-ice extent (SIE) stabilizes or recovers[13]. These major features of the CFC warming model are generally consistent with observations reported in our recent studies[7, 13].

Some $CO_2$-based climate models concluded that we must reduce carbon emissions to zero by 2050 to avoid more than 1.5 °C warming[14]. Other than $CO_2$, $N_2O$, and $CH_4$, there are certainly other well-mixed greenhouse gases (WMGHGs), particularly halo-GHGs, and short-term radiatively active agents such as aerosols and ozone that are altered due to air pollution[1]. The modeled radiative forcing of $CO_2$ exhibits a continuously positive trend (see Figure 1a), whereas that of halo-GHGs has shown a plateau since around 2000 due to the regulations by the Montreal Protocol and its Amendments (Figure 1b). Satellite- and ground-based data indicate that aerosol loadings exhibit predominantly negative trends since 2000 over NH mid-latitudes and SH continents, while increasing over South America (and South Asia and East Africa as well) (Figure 1c). These changes have led to a globally decreasing aerosol abundance, as assessed in the IPCC AR6[1]. The AR6 also states that tropospheric ozone increased by 30–70% across the NH from the mid-20th century to the mid-1990s, and that since then, free tropospheric ozone has increased by 2–7% per decade in most regions of NH mid-latitudes, 2–12% per decade in the sampled regions of the NH and SH tropics, and less than 5% per decade at SH mid-latitudes. Simulations by some climate models argued that these shorter-term radiative forcings likely cancel out, and warming is pretty much controlled by just $CO_2$ emissions (see, e.g., the IPCC Special Report on 1.5 °C warming[14]). This notion implies that surface warming stops when $CO_2$ emissions stop, i.e., that there is no lag in surface warming once $CO_2$ emissions reach zero[14]. This is the basis of the concept of the so-called "carbon budget". However, the simulated results by 'state-of-the-art' climate models in the IPCC AR6[1] show that although the short-term effective radiative forcings (ERFs) of aerosols and ozone essentially canceled out during the period 1970-2010, they have led to a positive net forcing of 0.4-0.5 W/m$^2$ from around 2010 up to the present (the green solid curve in Figure 1d). This gives rise to complexities in surface temperature change in the regions (e.g., mid-latitudes) that are experiencing drastic changes in level of air pollution, adding to the greenhouse effect of WMGHGs. For example, in a study of regional mean changes in surface air temperatures due to changes in GHGs, aerosol emissions and tropospheric $O_3$ levels, Wang et al.[15] showed that in 2050 under a carbon neutrality pathway, warming caused by aerosol reductions would dominate climate change over the (sub)regions with surface air temperature increases by 0.5–1.4 °C, much higher than the GHG-caused surface temperature increases of ≤0.2 °C and the tropospheric $O_3$ caused surface temperature decreases of ≤0.2 °C. Although the uncertainties in simulated ERFs of aerosols and ozone were assessed as the largest contributors to the overall ERF uncertainty since 1750 in the IPCC AR6[1], these ERFs should have been better constrained by direct observations of aerosol loadings and tropospheric $O_3$ since the mid-1990s. Moreover, there are pollution-free regions without such complexities, particularly the Antarctic that has had nearly zero air pollution since 1750 and the Arctic that had eliminated the pollution by around 2000 (see Figure 1c).

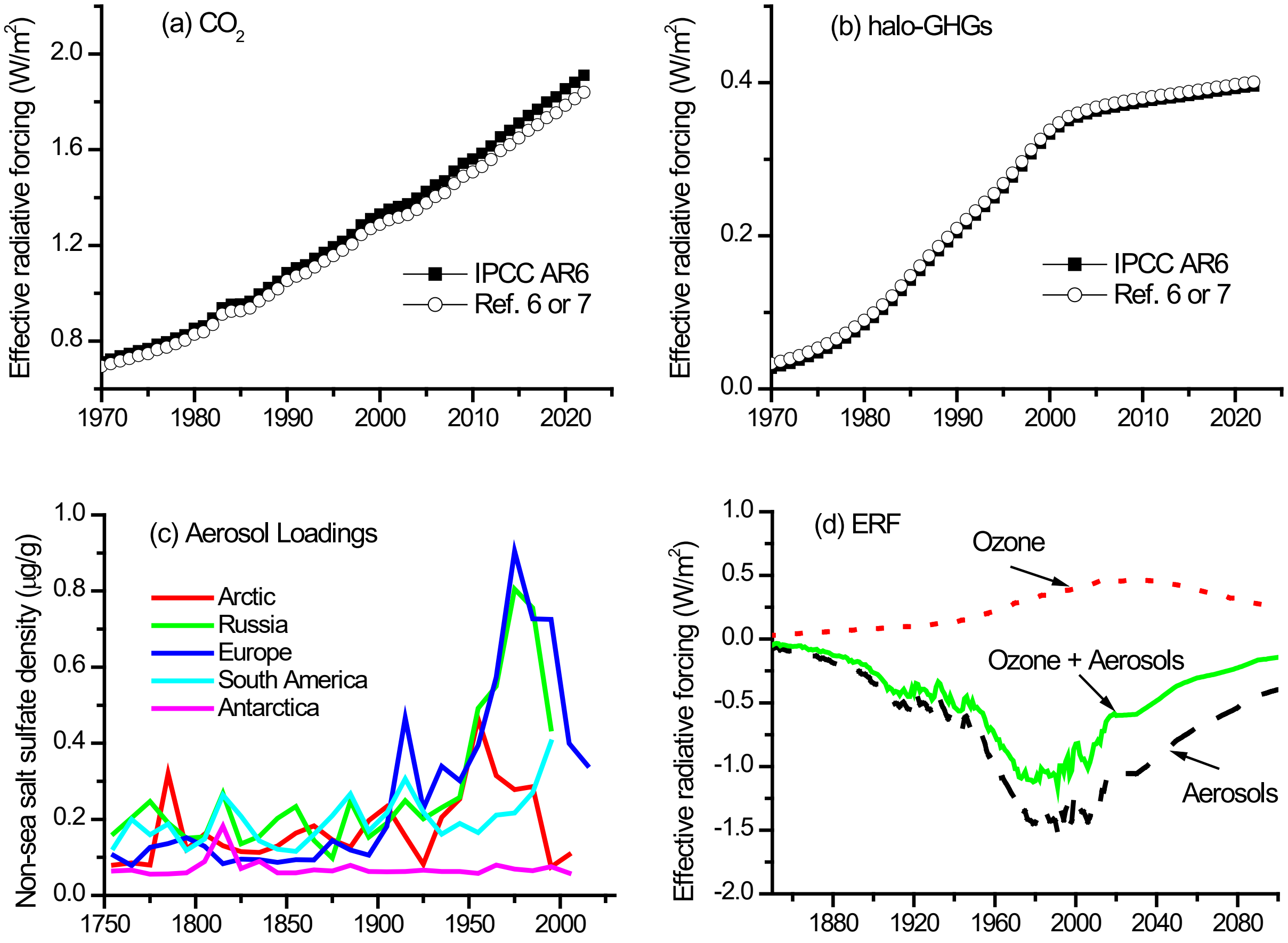

**Figure 1**. (**a-b**) Time-series effective radiative forcings (ERFs) of $CO_2$ and halo-GHGs; (**c**) 10-year averaged time series of aerosol loadings (concentrations of non-sea salt sulphate) from ice-core measurements; (d) time-series ERFs of (tropospheric and stratospheric) ozone, aerosols, and the net sum of ozone and aerosols, relative to the pre-industrial period in 1750. All of the data in 1a, 1b, 1c, and 1d, except the ERFs calculated in ref. [6] or ref. [7] in 1a and 1b, are obtained directly from the IPCC AR6 [1], in which ERFs were computed by GCMs for the future projection SSP245.

Additionally, there exist other climate drivers such as solar constant and surface albedo, as well as El Niño — Southern Oscillation (ENSO). ENSO is the most prominent year-to-year climate fluctuation on Earth, exerting its global impacts through atmospheric teleconnections, affecting extreme weather events worldwide[16, 17]. The total solar irradiance (TSI) relative to the mean in 1950-1970 had a rising trend from 1700 to 1950, but it has shown a slightly declining trend of $-0.7(\pm 1.7)\times10^{-3}$ (W/m$^2$)/decade since 1950, as shown in Figure 2a. Thus, solar constant variability has played a negligible role in surface warming since 1950 [1, 3, 4, 6, 10]. Surface albedo is mainly affected by SIE and snow-cover extent (SCE). There was a well-known amplification in Arctic warming, known as AA, which is a ubiquitous feature of the response to GHG forcing simulated by climate models [1]. It has been observed over recent decades that AA is closely related to Arctic sea ice loss [1, 18-21]. This relationship is also well observed in our recent studies[7, 13]. Observed data have shown that SCE in NH or North America has stabilized or shown a slight increasing trend of $0.9(\pm 1.7)\times10^4$ km$^2$/yr since the mid-1990s[13] (Figure 2b), and that SIE in NH largely decreased in 1995-2015 and since 2016, it has exhibited an increasing trend of 0.03(± 0.01) $km^2$/yr that was not reported in previous studies[1, 13], given by a linear fit to the observed data up to November 2024 (Figure 2c), while SH SIE had a slightly increasing trend in 1995-2015, suddenly dropped in 2015-2016, and since then it has had no significant trend (Figure 2d). As a result, the surface albedo effect and the associated AA effect on surface warming (especially over the NH), which were drastic during 1970-2015, are expected to have terminated since 2016. Thus, the polar regions are now free of both the incontrovertible large uncertainties in simulated ERFs of aerosols given by current climate models and the complexity caused by the AA effect. Therefore, the polar regions have now become an ideal "well-controlled laboratory" that can directly observe the climate effects of WMGHGs and ODS-caused ozone depletion, both of which are well monitored[7]. The data of all the potential climate drivers are updated and shown in Figures 1a-d and 2a-d, which are generally consistent with those presented in the IPCC AR6[1] and our recent studies[7, 13]. These data provide the basis for understanding climate changes from any climate model.

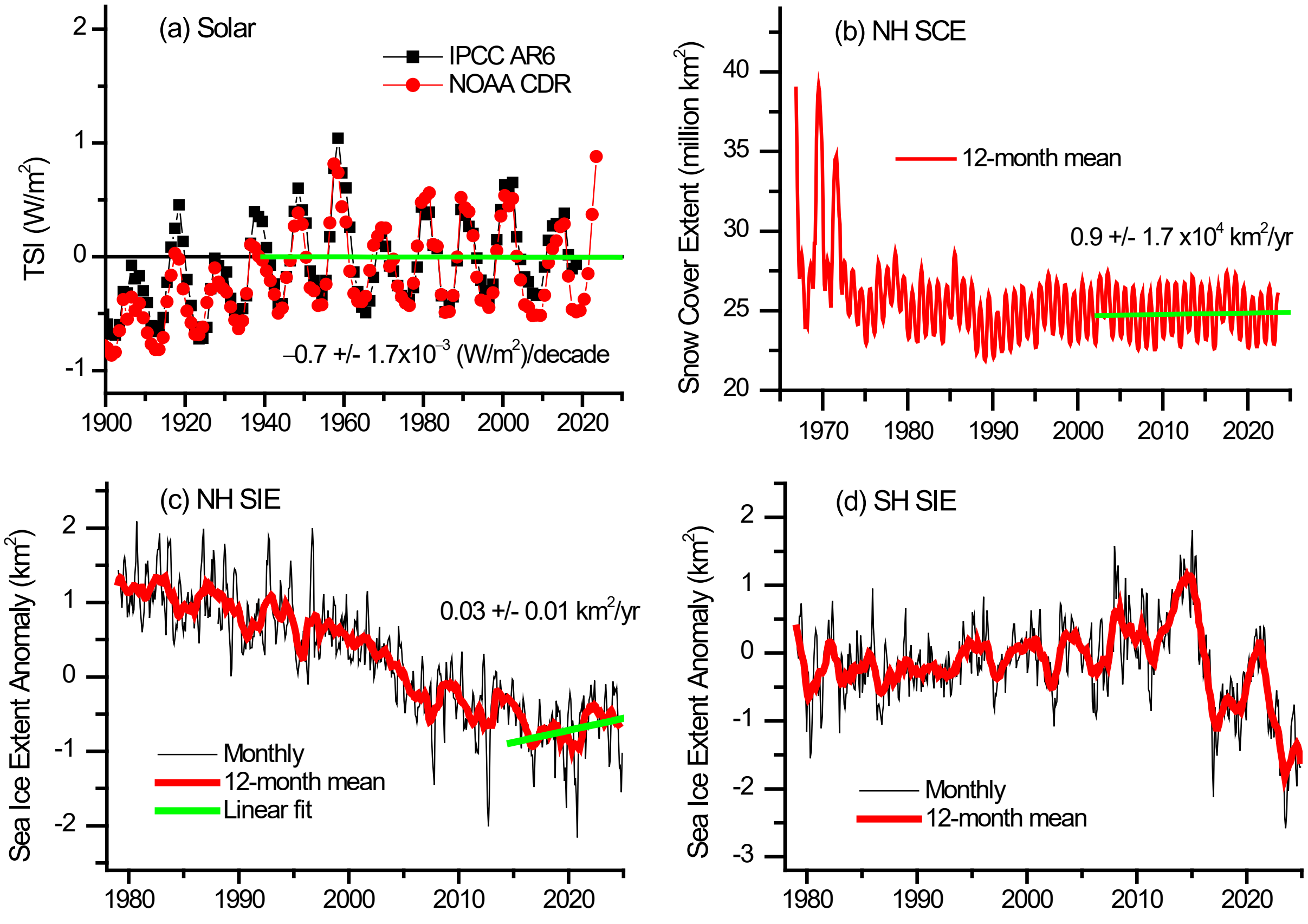

**Figure 2**. (**a-d**) Time series total solar irradiance (TSI) (relative to the 1950-1970 mean) in 1900-2023, NH snow cover extent (SCE) in 1968-2023, NH and SH sea ice extent (SIE) in 1979-(November) 2024. In (**a-c**), a linear fit to the observed data is given with the obtained trend indicated (the green line). All the data are obtained from the datasets of NOAA, together with the TSI data from IPCC AR6 [1].

Another key observation lies in changes in stratospheric temperature, which reflect changes in surface temperature[22-25]. In conventional climate models, changes in lower stratospheric temperature (LST) are primarily caused by ozone depletion due to emissions of (halogen-containing) ozone-depleting substances (ODSs)[26-28], whereas changes in upper stratospheric temperature (UST) are caused by changing concentrations of WMGHGs (mainly non-halogen GHGs, i.e., $CO_2$) and ozone depletion[22, 23]. $CO_2$-based climate models have also made an iconic prediction that tropospheric warming and stratospheric cooling would inevitably continue due to the projected continuous increases of $CO_2$ in the coming decades[1]. Indeed, radiosonde- and satellite-measured data showed that the troposphere warmed and the stratosphere cooled from the mid-1970s to the late 1990s[26-28]. However, most satellite measurements reviewed in the WMO report [26] and the IPCC AR6 [1] showed that the global LST has stabilized since ~1995. It is generally believed that the global LST has dominantly been controlled by ODSs[6, 8, 10, 11, 13, 26, 28]. It is noteworthy that the Manabe-Wetherald climate models[22, 23], recognized by the 2021 Nobel Prize in Physics, have the following important features. Generally, the sensitivity of the stratospheric equilibrium temperature to the change of GHG content such as $CO_2$ and water vapor is much larger than that of the surface equilibrium temperature, whereas the opposite is true for both cloudiness and surface albedo between the sensitivities of surface and stratospheric temperatures to them. Namely, the sensitivity of the equilibrium UST at 40 km to a change in GHG content is approximately five times that of the surface equilibrium temperature. In contrast, the influences of the surface albedo and cloudiness decrease with increasing altitude, having maxima at the earth's surface and being almost negligible at altitudes above 35 km. As a consequence, despite the dependences of temperatures extending from the surface to the lower stratosphere on multiple climate drivers including GHG content, ozone, surface albedo, and cloudiness that introduces the largest uncertainty in radiative forcing given by current climate models[1], the UST at altitudes above 35 km is predominantly dependent on the GHG content with relatively minor effects from ozone depletion or recovery. Given this key characteristic, time-series changes in UST if measured properly provide direct fingerprints of the change in total GHG effect. Fortunately, EUMETSAT's ROM SAF satellite data of the troposphere-stratosphere temperature climatology have been available since 2002, providing a source of observational information with high vertical resolution and long-term stability, covering from the troposphere to the upper stratosphere[29-31].

Compared with our earlier studies that used the observed data up to the late 2021/2022 [7, 13], three major advances and/or improvements are made in the current study. (i) We use the updated high-quality EUMETSAT's ROM SAF satellite datasets that have now covered the two complete solar cycles (the past 24 years: 2002-2023) and can therefore provide two 11-year average datasets for the periods 2002-2012 and 2013-2023 to minimize the solar cyclic effect for reliable UST trend analyses. Furthermore, time-series regional UST averaged at altitudes of 35-40 km over the Arctic (60°N-90°N), NH and SH mid-latitudes (30°N-60°N and 30°S-60°S), the tropics (30°S-30°N), and the Antarctic (60°S-90°S) will be presented. Moreover, multiple UST datasets from the EUMETSAT, NOAA, and the UK Met Office will be compared for reliable analyses of UST trends. (ii) More critically, a stabilizing or recovering trend of the sea-ice extent, which was not discovered in earlier studies [1, 13], is now observed for the first time in the monthly NH SIE data up to November 2024 (with an increment of about 24 data points), as shown in Figure 2c. As a result, the stopping of the associated AA effect on surface temperature in the Arctic is now expected to become detectable. (iii) Notably, no short-term radiative forcings of aerosols and ozone were included in our earlier calculations of GMST by the CFC-warming physics model[7, 13], which led to the somewhat lower GMSTs than the observed values after 2015. To improve our theoretical calculations of GMSTs, this study will re-calculate GMSTs to include all the ERFs of solar variability, halo-GHGs, aerosols and ozone and their individual contributions to GMST changes will also be presented. With these advances and improvements, this study aims to examine the predictions made by conventional climate models (GCMs) and the CFC-warming model and to precisely reveal the main cause of the GMST rise in the period 2015-2023. The results from this study may lead to the ending of the long-standing debates on the main cause of global warming since the 1970s [6, 8, 10, 32-36].

## 2. DATA AND METHODS

This study uses the following datasets: The radio occultation (RO) data directly provided by the Radio Occultation Meteorology Satellite Application Facility (ROM SAF) which is a decentralized operational RO processing center under EUMETSAT[37, 38]; The data of NOAA surface temperatures v6.0.0 [39], snow cover extent, sea ice extent, ENSO, total solar irradiance (TSI), and NOAA SSU/AMSU-A upper stratospheric temperatures (Channel 2), obtained directly from the NOAA; the data of TSI, concentrations of $CO_2$ and halogenated GHGs, effective radiative forcings of ozone and aerosols were obtained from the IPCC AR6 [1].

To compare with our theoretical GMST anomalies, the observed GMST data are presented with and without removal of the natural ENSO effect, which is the largest source of year-to-year variability[1, 16]. We simply adopt an empirical approach developed by Lean and Rind[3, 4] to remove the ENSO effect, with details given in previous studies[7, 13].

To calculate GMST anomalies over the last two decades, we use our CFC-warming physics model with details given previously[6, 7]. Here the CFC model[6, 7] is modified to take into account of radiative forcings (RFs) of not only halo-GHGs and solar variability but also ozone and aerosols, the GMST change ($\Delta T_s$) becoming

$$\Delta T_s = \lambda_c^{halo} \times (RF^{halo} + RF^{O3}) + \lambda_c^s \times (RF^{solar} + RF^{aerosol}), \qquad (1)$$

where $\lambda_c^s = 0.46 - 0.63\ K/(Wm^{-2})$ determined by an observational approach[5, 6] and $\lambda_c^{halo} = 1.77\ K/(Wm^{-2})$ determined from the energy spectrum of the Earth's blackbody radiation given by Planck's formula and the measured atmospheric transmittance spectrum[6]. Like halo-GHGs, ozone has a strong infrared absorption band in the atmospheric window at wavelengths of 8-13 μm and therefore has the same climate sensitivity to its forcing as halo-GHGs, while aerosols mainly interfere the solar energy reaching Earth's surface and hence share the solar climate sensitivity. RFs of halo-GHGs and solar variability are analytically calculated with a lag of 10 years for halo-GHG forcings[6, 7, 13]. Compared with similar approaches giving a linear equation of GMST anomalies expressed in terms of natural and anthropogenic forcings by Lean and Rind [3, 4] and Douglass and Clader [5], the CFC conceptual physics model uses no tunable (fitting) parameters. This is also in strong contrast to GCMs[1, 2]. Since this study aims to reveal the main cause of the change in GMST over the past two decades, all the forcings are relative to their respective values in 2000 (averaged over the period 1995-2005).

## 3. RESULTS and DISCUSSION

Figure 3 shows the results from the ROM SAF datasets for the 11-year mean temperature differences between the last two complete solar cycles (2002-2012 and 2013-2023) to minimize the effect of solar cycles. First, the results in Figures 3a-3b confirm that the overall warming pattern in the troposphere-lower stratosphere closely resembles the atmospheric distribution of CFCs. This observation provides strong visual evidence of the CFC-dominating mechanism of global warming, a conclusion reached previously [6, 7, 13] . There are clear cooling trends in the polar lower (10-25 km) and tropical lower-middle (22-35 km) stratospheres, which are mainly controlled by ozone depletion[26-28]. Note that although the levels of major ODSs (CFC-12 and CFC-11) are declining, the CR flux has been increasing over the past four solar cycles, leading to complex phenomena in ozone depletion[6, 10, 11]. To obtain a precise understanding of these LST changes, quantitative calculations of stratospheric ozone depletion will be needed and are being undertaken by the author. Second, most remarkably, the UST at altitudes of 35-40 km has been increased by 0.7 and 1.2 K between the last two solar cycles over the Antarctic and Arctic respectively, and warming trends at altitudes of 35-40 km over non-polar regions are about to emerge (Figures 3a and 3c). This leads to a warming trend in global mean UST (Figure 3c). As stated above and according to the Manabe-Wetherald climate models [22, 23] (see Figure 3d), this warming in UST provides the fingerprints that the total greenhouse effect of WMGHGs (i.e., the total GHG content) in the globe is decreasing, particularly over polar (high-latitude) regions. This critical observation is also pronounced from time-series data of mean UST at altitudes of 35-40 km available for the period 2002-2023 of the Arctic (60°N-90°N), NH and SH mid-latitudes (30°N-60°N and 30°S-60°S), the tropics (30°S-30°N), and the Antarctic (60°S-90°S), as shown in Figures 4a-e. Linear fits to the observed UST data show clear increasing trends of 0.77(±0.57) K/decade in the Arctic and of 0.69(±0.22) K/decade in the Antarctic, and no statistically significant trends of 0.01(±0.09) K/decade in the tropics and of −0.01(±0.17) K/decade / −0.31(±0.27) K/decade at NH/SH mid-latitudes. These results in Figures 3a, 3c and 4a-e together indicate that the total GHG effect has been significantly decreasing in polar regions and has not significantly changed in non-polar regions since the turn of the century. Given the well-measured increasing annual growth rates of atmospheric $CO_2$ in the past two decades [1], this observation sharply contradicts the prediction of enhanced cooling trends in UST by $CO_2$-based climate models, whereas it is in good agreement with the CFC warming model [6-8, 10, 13] (see Figures 1a-b).

Correspondingly, time-series data of surface temperatures for the period 2000-(November)2024 of the Arctic (60°N-90°N), Northern and Southern mid-latitudes (30°N-60°N and 30°S-60°S), the tropics (30°S-30°N), and the Antarctic (60°S-90°S), obtained from the most updated NOAA6.0.0 datasets [39], are shown in Figures 5a-e. It is particularly interesting to observe that consistent with the UST trends observed in Figures 4a-e, the surface temperatures exhibit a decreasing trend of −0.04(±0.06) K/decade in the Antarctic since 2005 and of −0.41(±0.37) K/decade in the Arctic since 2016, and a non-statistically significant (weak) warming trend of 0.14(±0.08) K/decade in the tropics. The latter was affected by recent warming spikes influenced by a rare 'triple-dip' La Niña. In striking contrast, there remain statistically significant increasing trends in surface temperature at both SH and NH mid-latitudes, which are 0.22±0.04 °C/decade and 0.78 ±0.12 °C/decade respectively, despite no significantly cooling trends in UST over these regions (Figures 4b and 4d). As acknowledged in the IPCC AR6 (Chapter 7, p. 987)[1] and demonstrated in our recent study[7], the observed changes in the Antarctic surface temperature since 1850 are drastically different from the equilibrium warming pattern simulated by conventional climate models[22, 23], which predicted an increase of 6-10 K for a $CO_2$ doubling, or by Earth system models (ESMs) under $CO_2$ radiative forcing even taking into account of ocean dynamics. In strong contrast, the measured surface temperatures in the Antarctic are well reproduced by the calculations using the CFC-warming model with inputs of halo-GHGs and ozone depletion alone [7]. Most remarkably, the Arctic surface temperature has a sensitive response to the emerging recovery in NH SIE (Figure 2c), consistent with the well-known dominance of the AA effect prior to 2016 [7, 13, 18-21]. Surface warming at mid-latitudes is consistent with the observed declining trends in atmospheric aerosol loading in the regions. As seen in Figure 1c, aerosol loadings are rapidly decreasing at mid-latitudes, which have negligible effects on UST at altitudes above 35 km[22, 23] but generate a positive radiative forcing (Figure 1d) leading to surface warming in these regions. In addition, there is also a small positive radiative forcing of ozone (Figure 1d). The results in Figures 4 and 5 clearly indicate that surface warming at mid-latitudes that take up 36.6% in area of the globe is caused by reduced air pollution and is responsible for the rise in GMST in the period 2016-2023.

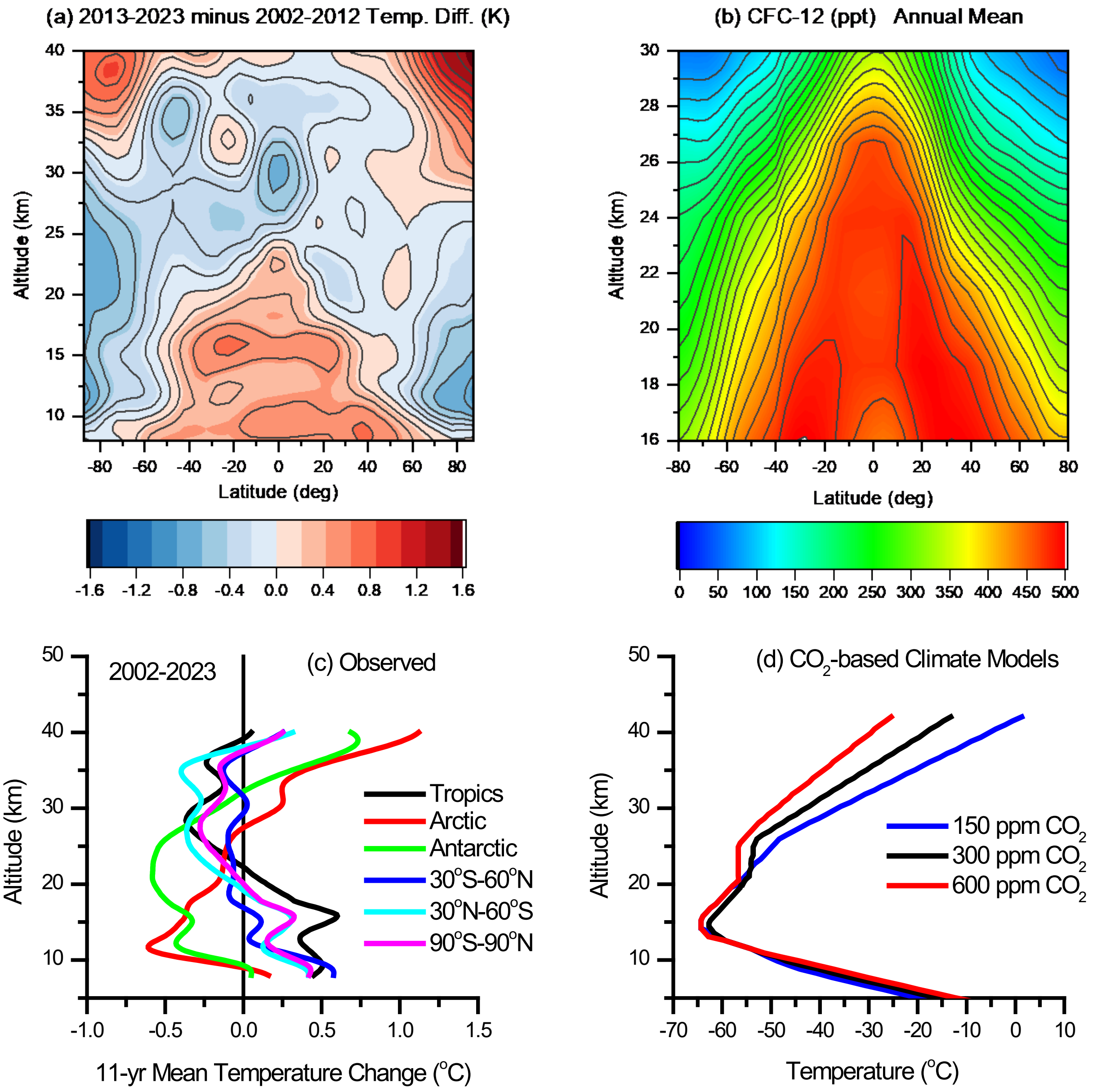

**Figure 3**. Tropospheric-stratospheric temperature (T) climatology, CFC spatial distribution, and $CO_2$-climate model prediction. (**a**): T difference in 11-yr mean zonal mean latitude–altitude distribution of the T climatology at altitudes of 8–40 km of 2013-2023 minus 2002-2012, obtained from the EUMETSAT's ROM SAF satellite datasets. (**b**): Representative zonal mean latitude–altitude distribution of the $CF_2Cl_2$ concentration, obtained from the NASA UARS's CLEAS dataset. (**c**): Observed altitude profiles of the 11-yr mean T difference over the last two solar cycles for the tropics (30° S–30° N), Antarctic (60°–90° S), Arctic (60°–90° N), mid-latitudes (30°–60° S/N) and the globe (90°S–90°N) of the Earth. (**d**): GCM-modeled altitude profiles of the variation in the troposphere and stratosphere T due to increased levels (150 ppm, 300 ppm, and 600 ppm) of atmospheric $CO_2$. (**a–c**): Modified and updated from Lu[7, 13]; (**d**): re-plotted from the Illustration and Popular science background for the 2021 Nobel Prize in Physics by the Royal Swedish Academy of Sciences, which originated from the Manabe-Wetherald climate model [22].

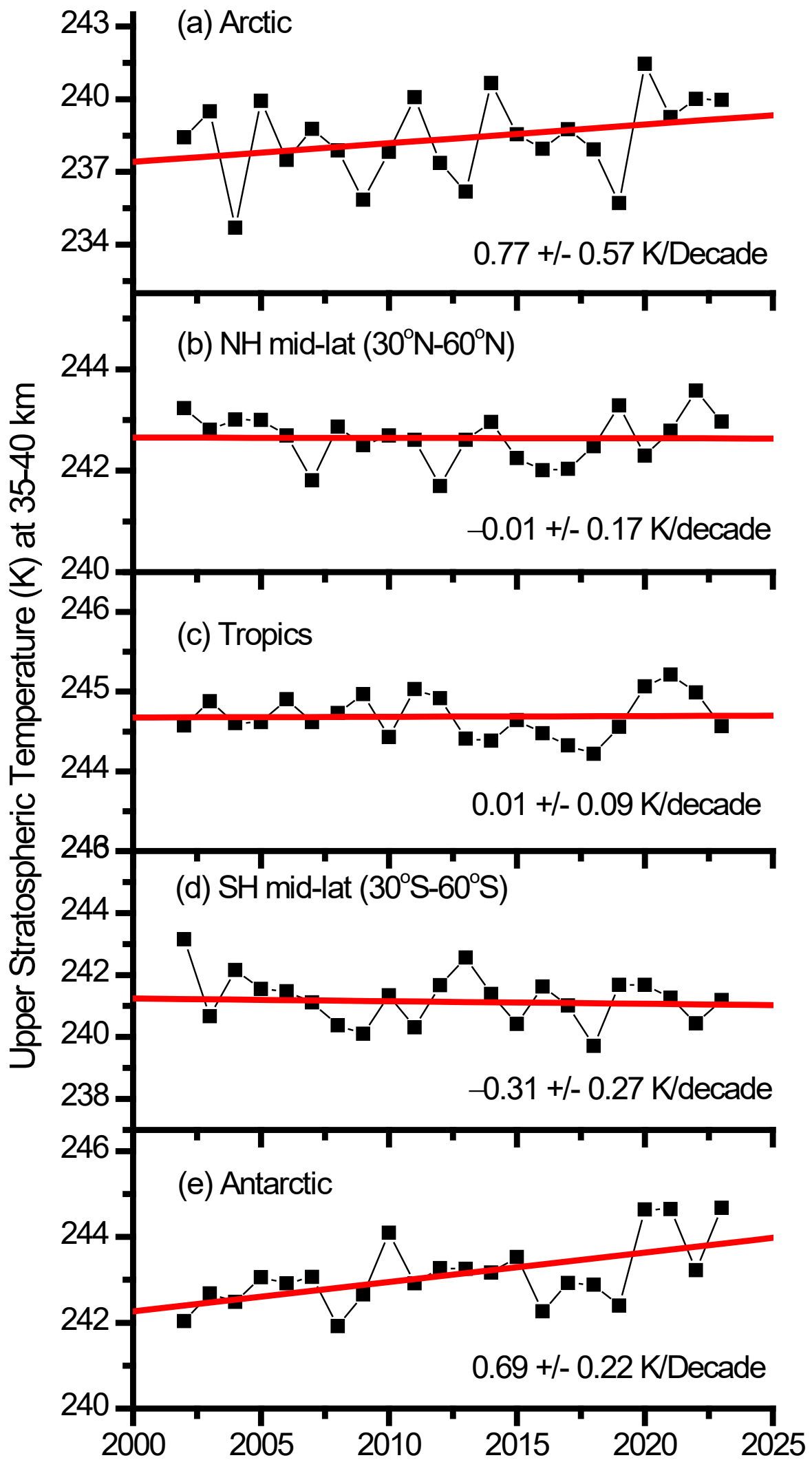

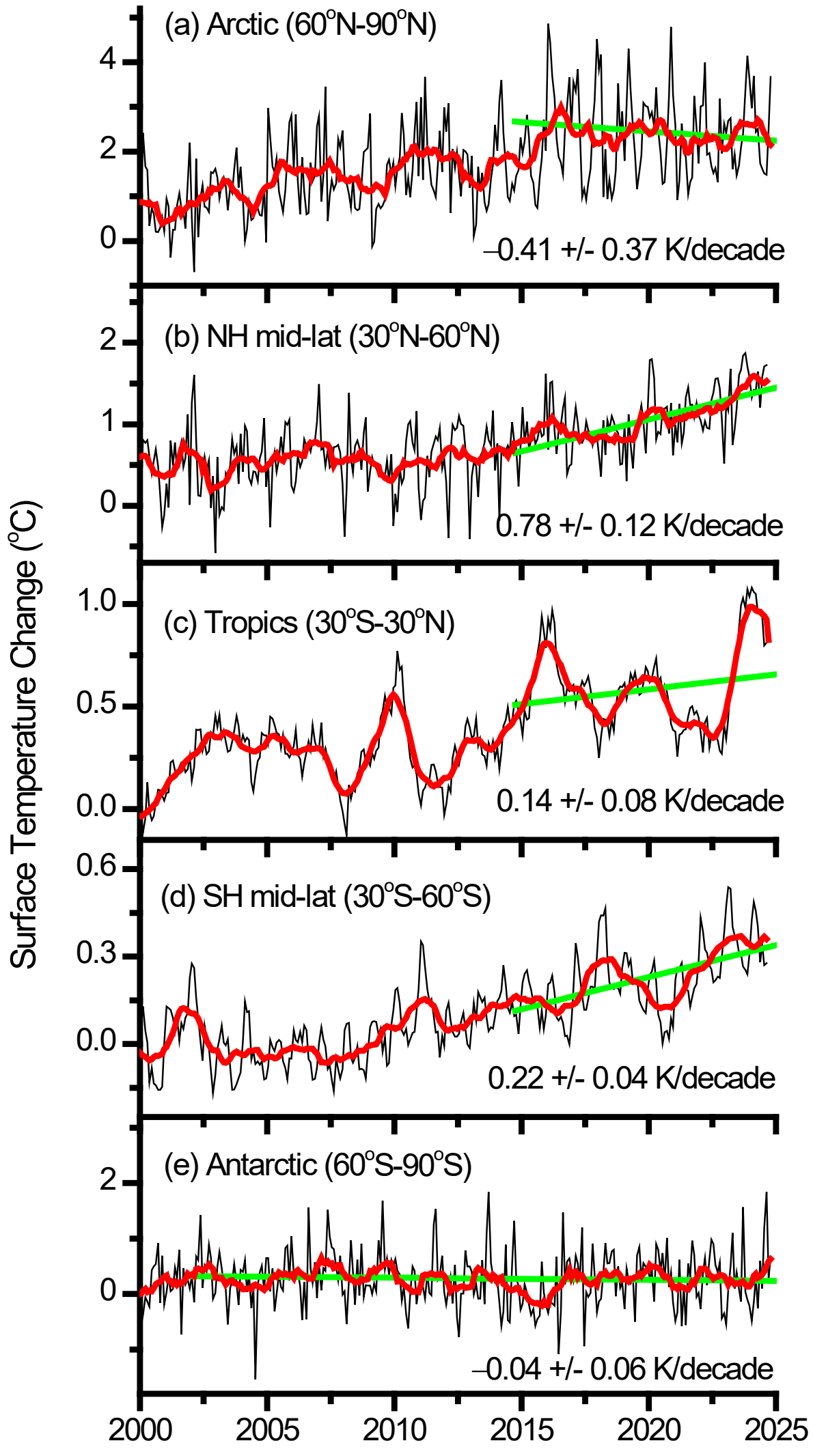

**Figure 4 (Left Panel)**. (a)-(e): Time-series annual upper stratospheric temperature (UST) averaged at altitudes of 35-40 km in the period 2000-2023, obtained from the EUMETSAT's ROM SAF satellite datasets. The red lines are the linear fits to the observed datasets with the slope ± its error indicated for each fit.

**Figure 5 (Right Panel)**. (a)-(e): Observed monthly regional surface temperatures up to November 2024 obtained from the NOAA 6.0.0. The red curves are the 12-month averaging to the observed datasets; the green lines are the linear fits to the observed datasets with the slope ± its error indicated for each fit.

To obtain a quantitative understanding of the GMST changes over the last two decades, we calculate GMST changes by the CFC-warming physics model with the details given in Data and Methods, in view of the fact that the modeled radiative forcings of $CO_2$, $CH_4$ and $N_2O$ are apparently missing in the observed data shown in Figures 3-5 and previously [6-8, 10, 13, 32-34, 40, 41]. As shown in Figure 6, our calculated results of GMST exhibit overall good agreement with the observed GMST data in 2000-2024. Remarkably, the calculated results given by the no-parameter physics model captures 76% ($R^2$) of the variance in the observed GMSTs. This value is exactly identical to that found by Lean and Rind [3, 4], though they used an empirical model including five fitted parameters (equivalently tunable climate sensitivities). It is demonstrated in Figure 6 that the CFC physics model closely reproduces the observed GMSTs from 2000 to 2024, including the almost no warming during 2000-2012 and the significant warming by 0.2-0.3 °C over the past decade. The observed data are presented both in original form and with removal of the contribution of the natural ENSO [3, 4, 7, 13]. Both the observed and calculated results of GMST show that the GMST was nearly constant during the first decade of this century (till about 2012), which was widely reported in the literature, as we discussed previously [7, 13]. We note that our calculated GMSTs after 2016 are slightly lower than the GMSTs given by Version 6.0.0 of the NOAA dataset. These small discrepancies, however, are more likely caused artificially by the NOAA data processing than by our CFC-warming model, since historical GMSTs in the period 2000-2012 given in the IPCC AR6[1] are higher by about 0.09 °C than those given in the IPCC AR5[27] due to 'changes in observational understanding' and we have demonstrated that GMSTs after 2016 in the version of either NOAA 5.1 or UK Met Office's HadCRUT5 are higher than those in their earlier versions (e.g., HadCRUT4.6 or NOAA 4.0) [7]. The four individual contributions of solar irradiance variability, halo-GHGs, ozone, and aerosols to GMST changes, are plotted in Figures 7a-7d, while Figure 7e shows the contribution of +0.27 °C from the net radiative forcing of aerosols and ozone to GMST changes during 2000-2024. All the results are relative to their respective 1995-2005 means. These results clearly show that the rise in GMST in the past decade is mainly caused by the positive short-term radiative forcings of ozone and aerosols as a result of reduced air pollution, which has occurred in mid-latitude regions, with relatively minor contributions from halo-GHGs and solar variability. This is consistent with the observed results of regional USTs and surface temperatures shown in Figures 4 and 5.

With the observations of rapidly lowered aerosol loading at (especially NH) mid-latitudes, slowly peaking halo-GHGs, and stopped Arctic amplification, we expect to see an emerging reversal in GMST. It is worth noting that the current net radiative forcing from ozone and aerosols has returned and been close to its global level in the 1950s (Figure 1d). Thus, the effects of aerosols and ozone on GMST in the coming decades should be very limited. Moreover, the TSI peak in the current solar cycle and the ENSO warming peak occurred in October and November 2023 respectively, as seen in observed data, and the responses in GMST to TSI and ENSO variability had short delays of only 1-4 months [3, 4, 13]. In fact, both the observed GMST data, with or without removal of the ENSO effect using the established approach by Lean and Rind [3, 4], exhibit a peak around October 2023 and since then, it has been in a decreasing trend over the past 13-14 months. This observation is particularly interesting in that the TSI is peaking in the current solay cycle and an expected shift to cooling La Niña pattern in the Pacific Ocean has been delayed, which is now projected to emerge weakly in December 2024 and is likely to limit the cooling influence of the climate pattern on global average temperatures. Thus, we are optimistic to observe a continuous long-term reversal in GMST, which likely already started in the end of 2023 and will exhibit a cooling trend in the coming decades (Figure 6).

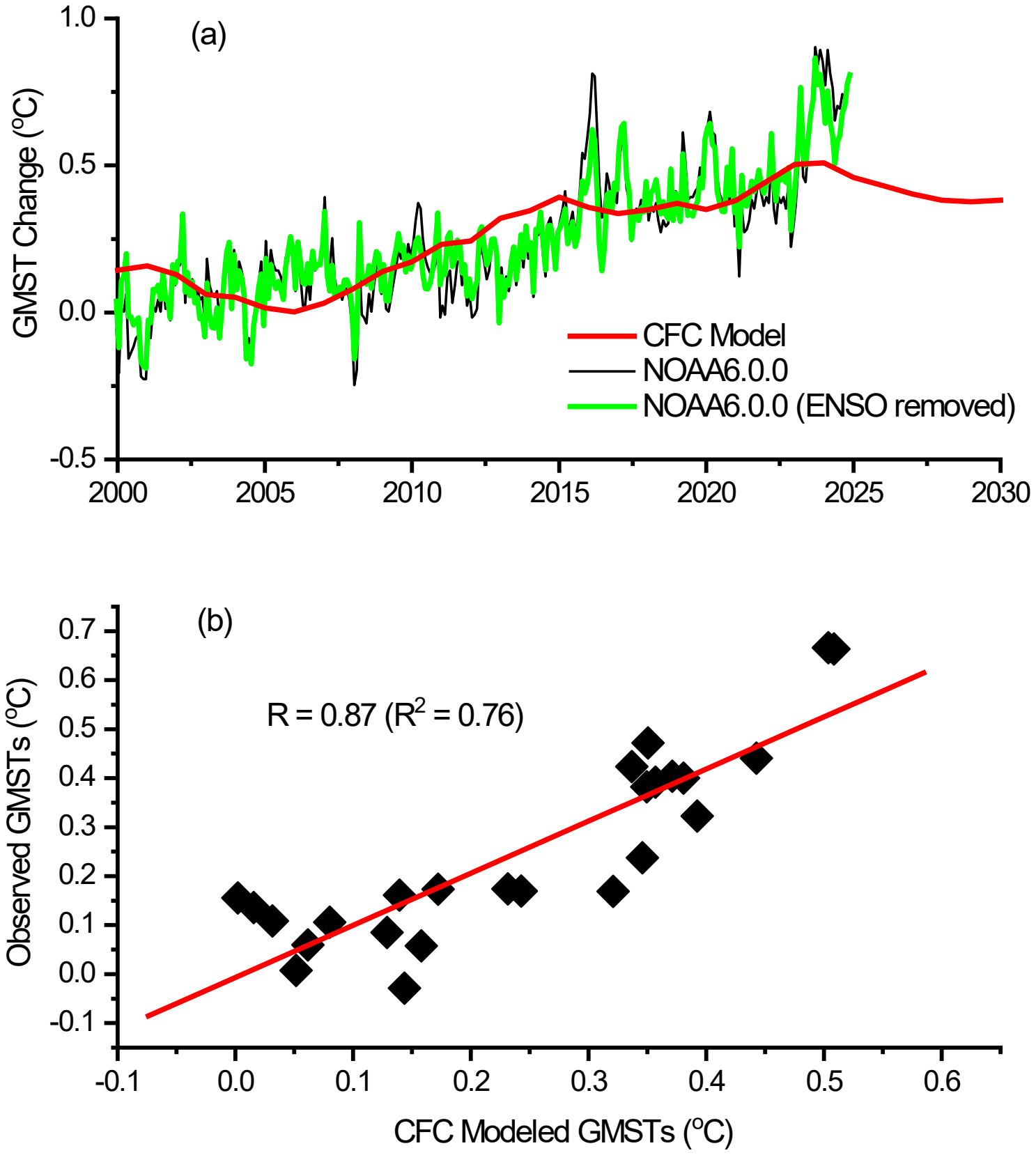

**Figure 6**. **(a)** Observed and calculated GMSTs. Observed GMST data were the NOAA6.0.0 combined land surface air temperature and sea surface temperature anomalies up to November 2024 (thin black line: original monthly data; thick green line: after removal of the natural ENSO effect). Calculated GMSTs (thick red line) are obtained by the CFC-warming model including the contributions of halo-GHGs, the solar effect, ozone, and aerosols. All the results are relative to their 1995-2005 means. **(b)** A statistical analysis gives a correlation coefficient of R = 0.87 ($R^2 = 0.76$) for the observed GMSTs and no-parameter CFC model, indicating that the theoretical results account for 76% of the variance in the observed GMSTs from 2000 to 2024.

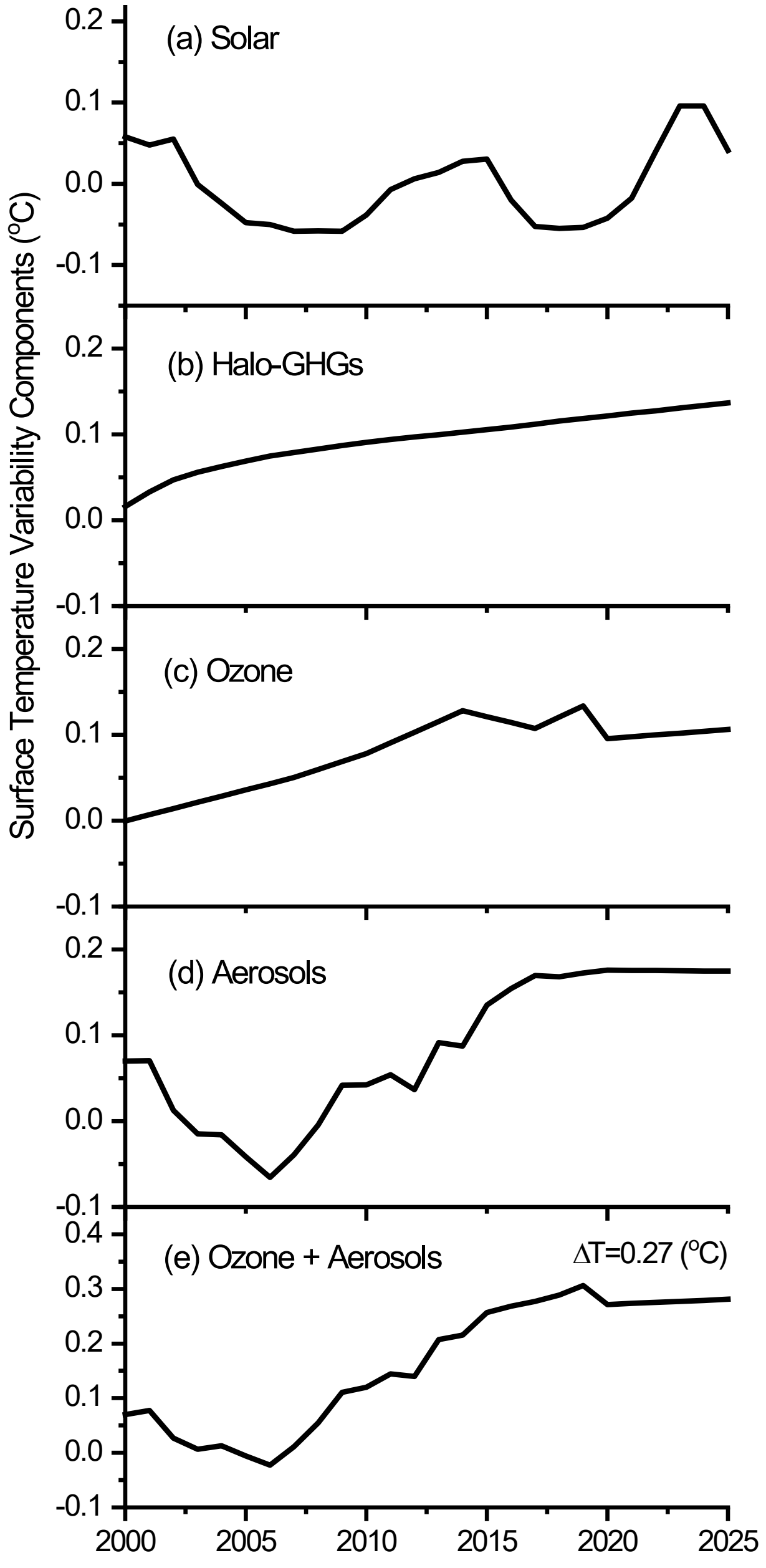

**Figure 7.** (a-d) Calculated individual contributions of the solar effect, halo-GHGs, ozone, and aerosols and (e) the sum contribution of ozone and aerosols to the GMST change by the CFC-warming physics model. Each of the results is relative to the 1995-2005 mean.

Finally, we must note that our observed UST trends in ROM SAF data displayed in Figures 3 and 4 show different trends from the NOAA STAR reprocessed UST data of Stratospheric Sounding Unit (SSU) and Advanced Microwave Sounding Unit (AMSU)-A (Channels 1-3). The latter provide data on temperatures in the middle and upper stratosphere, but they encounter formidable problems in deriving UST trends, which are subject to calibration issues in the SSU/AMSU space-view anomaly and radiometric anomalies, changing pressures in the pressure modulator cells, solar tides, and atmospheric $CO_2$ concentration changes in the satellite observations [24, 25, 42, 43]. In the two independently processed SSU datasets [24, 25, 42, 44], the NOAA STAR data were found to have striking differences in stratospheric temperature trends from those earlier processed by the UK Met Office (UKMO) [45]. For example, the global-mean cooling in SSU channels 1 and 2 sampling the middle stratosphere (around 25–45 km) in the NOAA data set was nearly twice as large as that in the UKMO data set. The differences between the NOAA and UKMO global-mean time series were so large that they called into question the fundamental understanding of observed temperature trends in the middle and upper stratosphere from the SSU data [45]. Since the SSU measured the radiances from emission by atmospheric $CO_2$, the weighting functions of its different channels were subject to time-varying changes, which was based on the understanding from $CO_2$ climate models[42, 44]. Owing to increasing atmospheric $CO_2$, the weighting functions of the three SSU channels were modeled to shift to higher altitudes at higher temperatures, and hence spurious positive temperature trends were presumably superimposed on the SSU temperature trends and corrected from the original data in the NOAA STAR dataset [42, 44]. Not surprisingly, such revised trends become significantly more negative (cooling) in the middle and upper stratosphere, though they are expectedly in better agreement with $CO_2$ model-calculated trends that are expected to be negative throughout the stratosphere. However, the simulated impact of changes in $CO_2$ on the weighting functions and hence the measured radiances can be seriously problematic, provided with the documented satellite observations showing that the 15-μm atmospheric $CO_2$ absorption band was actually missing in the observed radiance difference spectrum in the rapidly warming period 1970-1999 [6, 7, 40, 41]. The latter observation is in striking contrast to what is expected from $CO_2$-based climate models. Although both NOAA and UKMO SSU datasets have been reprocessed and documented and differences between them have been reduced, there remain major discrepancies [43, 46, 47].

As shown in Figure 8, the UKMO reprocessed dataset [43] for SSU2 (at altitudes of 35-45 km) seems consistent with the global mean UST trend at 35-40 km obtained from the ROM SAF data but the NOAA STAR SSU/AMSU-A dataset does not, though the UKMO data are available only as globally 6-month mean data. Slightly differently from the NOAA STAR reprocessing, the UKMO data reprocessing used the computed weighting functions for the SSU channels with a selected temperature anomaly profile to produce a temperature best matching the observed temperature changes in each SSU channel[43]. It is worth noting that the UKMO reprocessed SSU time series were very close to those analyzed by several independent groups after the recommended corrections for radiometric, spectroscopic and tidal differences to resolve the discrepancies among SSU datasets[43]. The consensus results were the following: a cooling trend over 1979–1996 was seen in all three SSU channels, while a weakly warming trend of 0.13±0.1 K in Channel 2 and no statistically significant cooling trends of −0.05±0.1 K in Channel 1 and −0.12±0.3 K in Channel 3 over 1997-2006 were observed[43]. Similarly, Randel at al.[48] found near zero trends in time-series de-seasonalized temperature anomalies for SSU2 at the equator, 50°S, and 80°S during 2000-2015, with data from the NOAA SSU, the Aura Microwave Limb Sounder (MLS) and Sounding of the Atmosphere Using Broadband Emission Radiometry (SABER) instruments. In Figure 8, a linear fit to the ROM SAF data gives a weakly warming trend of 0.04±0.03 K/decade during 2002-2023, which is very close to the weakly warming trend during 1997-2006 observed by Nash and Saunders[43] and the near zero trend during 2000-2015 observed by Randel at al.[48] Notably, the pronounced UST cooling trends over 2002-2023 in all three SSU channels obtained from the NOAA STAR data (Channels 1 and 3 are not shown) are not consistent with the observed trends in surface temperature (Figure 5), independent of $CO_2$-based climate models or the CFC-warming model. It seems very likely that the UST cooling trends derived from the model-corrected NOAA STAR datasets arise from artificial effects caused by the above-mentioned data calibration problems.

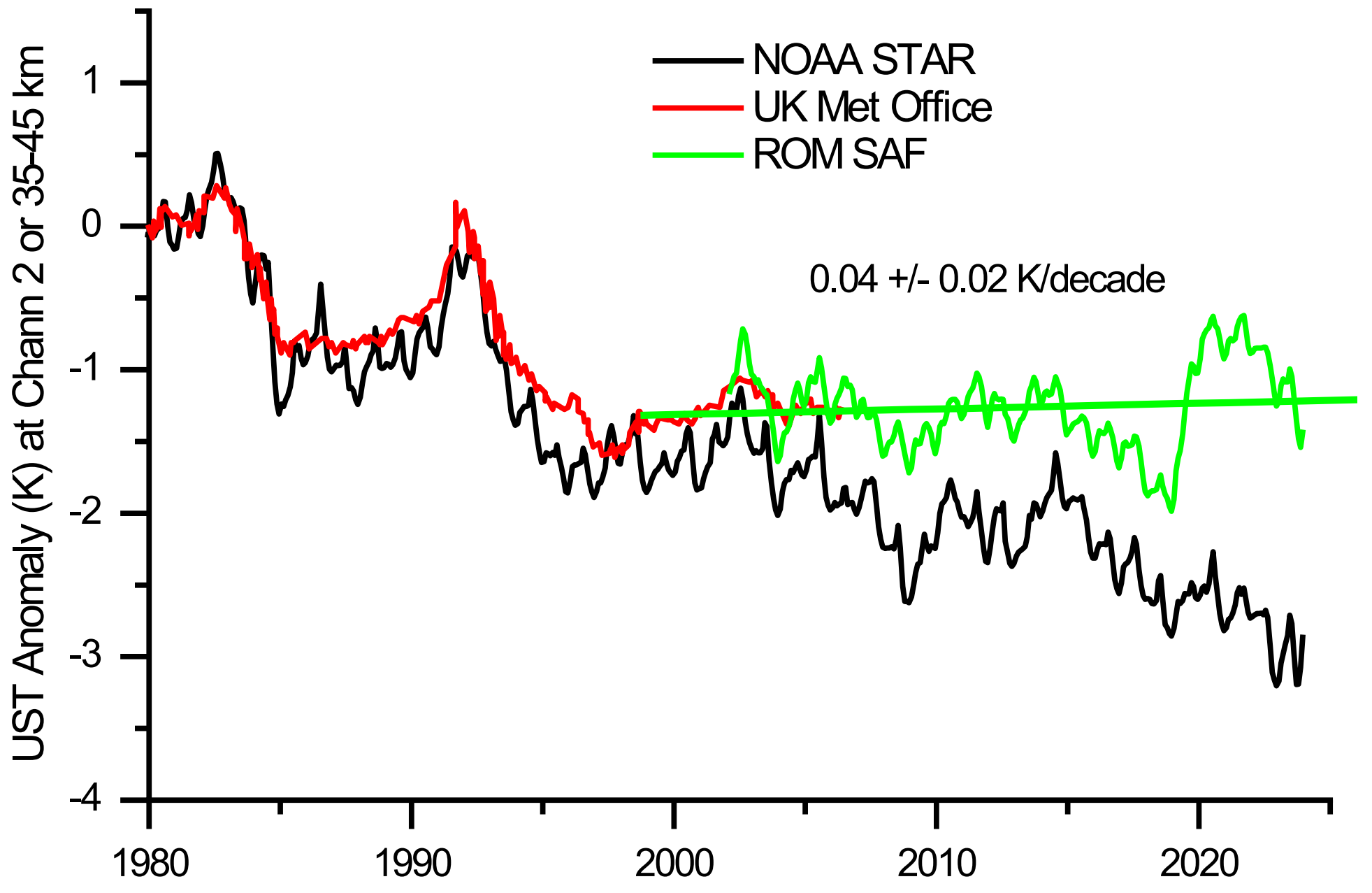

**Figure 8.** Time-series upper stratospheric temperatures (USTs) in 1980-2023, obtained from the EUMETSAT's ROM SAF satellite dataset at altitudes of 35-40 km, UK Met Office (UKMO) and NOAA STAR SSU/AMSU-A datasets (Channel 2, 35-45 km). The green line is a linear fit to the ROM SAF dataset with the slope ± its error indicated.

In the literature, Loeb et al.[49] recently made an observational assessment of changes in Earth's energy imbalance since 2000. Their satellite observations from the Clouds and the Earth's Radiant Energy System show that global mean net flux (NET) at the top-of-atmosphere (equivalently Earth's energy imbalance) has doubled during the first twenty years of this century. This doubling is due to a marked increase in absorbed solar radiation (ASR) that is partially offset by an increase in outgoing longwave radiation (OLR). They also showed that the increase in ASR is associated with large decreases in stratocumulus and middle clouds over the NH sub-tropics, decreases in low and middle clouds at NH midlatitudes, and decreases in SH middle cloud reflection. The decrease in cloud fraction and higher sea-surface temperatures over the NH sub-tropics lead to a significant increase in OLR from cloud-free regions. They found that despite marked differences in ASR and OLR trends during the hiatus (2000–2010), transition-to-El Niño (2010–2016) and post-El Niño (2016–2022) periods, trends in NET are rather stable (within 0.1 $Wm^{-2}$ per decade). The latter was interpreted to imply "a steady acceleration of climate warming". Their study called for new climate model simulations, which are critically needed to understand such climate change.

Tselioudis et al.[50] recently analyzed zonal mean cloud and radiation trends over the global oceans for the past 35 years from a suite of satellite datasets covering two periods (1984-2018 and 2000-2018). Their zonal mean oceanic total cloud cover (TCC) trend analyses showed positive or near-zero cloud cover trends in high-latitude (≥50° S or N) zones and negative trends in low-latitude zones (50°S-50°N). The latter dominate, leading to decreases in global mean TCC. These contrasting TCC changes between the high and low latitudes produce contrasting low-latitude cloud radiative warming and high-latitude cloud radiative cooling effects, present in both the ISCCP-FH and CERES-EBAF datasets. The balance between the two contrasting trends leads to the global ocean mean trend of the shortwave cloud radiative effect (SWCRE), which has the oppositive warming and cooling trends shown from the two datasets. When the period is extended from 2016 to 2022, the CERES cloud radiative warming trend doubles in magnitude, indicating a stronger cloud radiative heating in the past 6 years coming from the low latitudes.

A number of different mechanisms were proposed to explain the observed contrasting TCC trends in the high- and low-latitude regions[50]. These include the expansion of the low-TCC subtropical zone due to the changes in atmospheric general circulation (a tropical widening and a shift of the midlatitude storm tracks), local mechanisms rather than large scale shifts indicated by the observed TCC trends inside the core of the latitude zones, as well as indirect effects from aerosol decreasing trends. The tropical widening is suggested to arise from stratospheric ozone depletion, aerosol and ozone forcing, and also coupled atmosphere–ocean variability. They also suggested a local mechanism for the TCC increase at high latitudes, which is related to transitions from ice to water cloud particles as temperature warms which may result in both larger cloud amounts and larger cloud optical depths through the 'cloud phase feedback'. The TCC decrease at low latitudes that dominates the global TCC trend is suggested to be the result of cloud reduction mechanisms and indirect effects from aerosol decreasing trends.

Although the results from such cloud cover studies by Loeb et al.[49] and Tselioudis et al.[50] are very interesting and are not inconsistent with the positive radiative forcing as a result of reduced aerosols associated with improved air quality at sub-tropics and mid-latitudes used in our current calculations of GMSTs, the interpretations of those cloud cover observations may be rather challenging and need a strong support from quantitative estimates from a proper theory. For example, both cloud cover studies show a poor connection between the cloud cover and surface temperature trends during the hiatus (2000–2012) and the warming period 2013-2022 or during the periods 1984-2000 and 2000-2018 (2022) and in different regions. Particularly, the surface temperature trends in the Antarctic and Arctic regions were drastically different during 1980/2000-2015 (Figures 5a and 5e). These are not captured in the observed cloud cover data. Furthermore, recognized climate models have demonstrated that aerosols and clouds have little influence on the UST variations[22, 23]. As noted in the Introduction and shown in Figure 1c, moreover, the aerosol loadening in the Arctic drastically decreased during the period 1950-2000 and both the Antarctic and Arctic have been almost pollution free since 2000. Consistently, our current results show that the increasing trends in UST over the Antarctic and Arctic since around 2000 are very similar (Figure 4), so are the decreasing trends in the polar surface temperatures since 2016 when the SIE stabilized or started to recover (Figure 5). Thus, it seems robust to conclude that the changes in upper-stratospheric and surface temperatures in the polar regions during the past decade were mainly caused by the greenhouse effect of halo-GHGs rather than the changes in cloud cover or aerosols.

## 5. CONCLUSIONS

In contrast to the predicted continuous warming on the surface and cooling in the upper stratosphere by $CO_2$-based climate models, a consistent phenomenon observed from both UST and surface temperature measurements on the global scale has emerged. The observed data strikingly show that the UST at altitudes of 35-40 km exhibits warming trends in polar regions and no significant trends in non-polar regions since 2002, and that correspondingly surface temperature exhibits cooling trends in the Antarctic since 2005 and in the Arctic since 2016 when the SIE-associated AA effect stopped, and no statistically significant trend in the tropics since 2016. In contrast, there remain statistically significant surface warming trends at SH and NH mid-latitudes, which are responsible for the recent rise in GMST in the last two decades. This warming, however, is quantitatively explained by the IPCC-documented positive short-term radiative forcings of aerosols and ozone as a result of improved air quality. According to the well-recognized climate models, the observed UST trends provide strong fingerprints that the total GHG effect has been decreasing in polar regions and not significantly increased in non-polar regions in the last two decades, leading to an overall decreasing trend in global GHG effect. These observations are in quantitative agreement with the calculated results by the parameter-free physics model based on halo-GHGs, whose destruction is consistent with the features of the CRE mechanism. The latter gives larger rates in destruction of halo-GHGs and hence earlier reversals in surface and upper-stratospheric temperatures at higher latitudes (polar regions). With observations of rapidly lowered aerosol loading at mid-latitudes, projected halo-GHGs, and terminated AA effect, we expect to see an emerging reversal in GMST, which likely started in the end of 2023 and will show a cooling trend in the coming decades, as consistently shown previously with the inclusions of different halo-GHG species (CFCs, HCFCs, HFCs and PFCs)[6, 7, 10, 13] and in Figure 6 of this current study. If our CFC-warming physics model is proven to the dominant mechanism for global climate change since 1950, we will observe continuous and emerging warming trends in UST over the polar regions and non-polar regions respectively and correspondingly surface cooling trends that have been observed in the polar regions and will be observed over the mid-latitude regions first and finally over the tropical regions. Future observations will decide whether these predictions are true and which warming mechanism prevails.

This study also leads to a perspective, which can be considered as good news to the world. That is, with the phasing out of halo-GHGs (CFCs, HCFCs, HFCs and PFCs) by international Agreements, including the most successful and important Montreal Protocol and its Amendments, it is very likely to see a gradual reversal in GMST in the coming decades. Thus, this study emphasizes the importance of such efforts from international governments and global community. On the other hand, it is equally important to emphasize that the relevant international policies and political agenda must be built on a solid scientific foundation that captures the right culprit of global climate change. Another implication from our recent studies[7, 12] and this study is to enhance concerns about geoengineering that has been proposed as a potential method to reduce climate warming by increasing sunlight reflection through the intentional addition of aerosols into the stratosphere, known as 'stratospheric aerosol injection'. In view of both our substantial observations on ozone and climate changes and the significant enhancement (harmful) effect of atmospheric cloud and/or aerosol particles on global ozone depletion[12], this study reiterates the author's strong recommendation not to proceed with the geoengineering project that is intensely pushed to make a move and will likely prove to be completely unnecessary. The identifying of halo-GHGs as the primary anthropogenic climate driver may turn out to be of critical importance for humans not only to reverse the climate change but to maintain a healthy economy and ecosystem across the globe.

## ACKNOWLEDGEMENTS

The author is greatly indebted to the Science Teams for making the data used for this study available. The author also thanks Professor Anthony Leggett for his helpful and constructive comments on an earlier version of this manuscript. This research was funded by the Natural Science and Engineering Research Council of Canada and the University of Waterloo.

## AUTHOR DECLARATIONS

The author has no conflicts to disclose.

## DATA AVAILABILITY

The data used for this study were obtained from the following sources: The radio occultation data were provided by the Radio Occultation Meteorology Satellite Application Facility (ROM SAF) which is a decentralized operational RO processing center under EUMETSAT. ROM SAF RO data are available at: http://www.romsaf.org [37, 38]. The data of NOAA surface temperatures (v6.0.0) ( https://www.ncei.noaa.gov/products/land-based-station/noaa-global-temp) [39], snow cover extent and sea ice extent (https://www.ncei.noaa.gov/access/monitoring/snow-and-ice-extent/), ENSO (https://www.cpc.ncep.noaa.gov/data/indices/), total solar irradiance (TSI) (https://www.ncei.noaa.gov/products/climate-data-records/total-solar-irradiance), and NOAA SSU/AMSU-A upper stratospheric temperatures (Channel 2) were obtained from the NOAA (https://www.star.nesdis.noaa.gov/smcd/emb/mscat/); the data of TSI, concentrations of $CO_2$ and halogenated GHGs, effective radiative forcings of ozone and aerosols were obtained from the IPCC AR6 (https://www.ipcc.ch/report/ar6/wg1/) [1]; zonal mean latitude–altitude distribution of $CF_2Cl_2$ (CFC-12) was obtained from the NASA UARS (CLEAS) dataset (https://earthdata.nasa.gov/)[13].

## REFERENCES:

1. IPCC, AR6 Climate Change 2021: The Physical Science Basis; Cambridge University Press: Cambridge, UK, 2023.
2. V. Balaji, F. Couvreux, J. Deshayes, J. Gautrais, F. Hourdin, C. Rio, Are general circulation models obsolete? Proc. Nat. Acad. Sci. (PNAS), 119 (2022) e2202075119.

3. J.L. Lean, D.H. Rind, How natural and anthropogenic influences alter global and regional surface temperatures: 1889 to 2006, Geophys. Res. Lett., 35 (2008) L18701.
4. J.L. Lean, D.H. Rind, How will Earth's surface temperature change in future decades? Geophys. Res. Lett., 36 (2009) L15708.
5. D.H. Douglass, B.D. Clader, Climate sensitivity of the Earth to solar irradiance, Geophys. Res. Lett., 29 (2002) 33-31-33-34.
6. Q.-B. Lu, New Theories and Predictions on the Ozone Hole and Climate Change; World Scientific: Hackensack, NJ, USA, 2015; pp.1–285.
7. Q.-B. Lu, Critical Review on Radiative Forcing and Climate Models for Global Climate Change since 1970, Atmosphere, 14 (2023) 1232.
8. Q.-B. Lu, Cosmic-ray-driven electron-induced reactions of halogenated molecules adsorbed on ice surfaces: Implica-tions for atmospheric ozone depletion and global climate change, Phys. Rep., 487 (2010) 141-167.
9. Q.-B. Lu, What is the Major Culprit for Global Warming: CFCs or $CO_2$? J. Cosmol. 2010, 8, 1846–1862.
10. Q.-B. Lu, Cosmic-ray-driven reaction and greenhouse effect of halogenated molecules: culprits for atmospheric ozone depletion and global climate change, Int. J. Mod. Phys. B, 27 (2013) 1350073.
11. Q.-B. Lu, Fingerprints of the cosmic ray driven mechanism of the ozone hole, AIP Adv., 11 (2021) 115307.
12. Q.-B. Lu, Formulation of the cosmic ray–driven electron-induced reaction mechanism for quantitative understand-ing of global ozone depletion, Proc. Nat. Acad. Sci. (PNAS), 120 (2023) e2303048120.
13. Q.-B. Lu, Major Contribution of Halogenated Greenhouse Gases to Global Surface Temperature Change, Atmosphere, 13 (2022) 1419.
14. IPCC, Special Report on Global Warming of 1.5°C; Cambridge University Press: Cambridge, UK, 2018.
15. P. Wang, Y. Yang, D. Xue, L. Ren, J. Tang, L.R. Leung, H. Liao, Aerosols overtake greenhouse gases causing a warm-er climate and more weather extremes toward carbon neutrality, Nat. Comm., 14 (2023) 7257.
16. S.-W. Yeh, J.-S. Kug, B. Dewitte, M.-H. Kwon, B.P. Kirtman, F.-F. Jin, El Niño in a changing climate, Nature, 461 (2009) 511-514.
17. J.D. Rojo Hernández, Ó.J. Mesa, U. Lall, ENSO Dynamics, Trends, and Prediction Using Machine Learning, Weather and Forecasting, 35 (2020) 2061-2081.
18. J.A. Screen, I. Simmonds, The central role of diminishing sea ice in recent Arctic temperature amplification, Nature, 464 (2010) 1334-1337.
19. A. Dai, D. Luo, M. Song, J. Liu, Arctic amplification is caused by sea-ice loss under increasing CO2, Nature Comm., 10 (2019) 121.
20. E.-S. Chung, K.-J. Ha, A. Timmermann, M.F. Stuecker, T. Bodai, S.-K. Lee, Cold-Season Arctic Amplification Driven by Arctic Ocean-Mediated Seasonal Energy Transfer, Earth's Future, 9 (2021) e2020EF001898.
21. R. Davy, P. Griewank, Arctic amplification has already peaked, Environ. Res. Lett., 18 (2023) 084003.
22. S. Manabe, R.T. Wetherald, Thermal Equilibrium of the Atmosphere with a Given Distribution of Relative Humidity, J. Atmos. Sci., 24 (1967) 241-259.

23. S. Manabe, R.T. Wetherald, The Effects of Doubling the $CO_2$ Concentration on the climate of a General Circulation Model, J. Atmos. Sci., 32 (1975) 3-15.
24. J. Nash, G.F. Forrester, Long-term monitoring of stratospheric temperature trends using radiance measurements ob-tained by the TIROSN series of NOAA spacecraft, Adv. Space Res., 6 (1986) 37-44.
25. J. Nash, Extension of explicit radiance observations by the Stratospheric Sounding Unit into the lower stratosphere and lower mesosphere, Q. J. R. Meteorol. Soc., 114 (1988) 1153-1171.
26. WMO/UNEP, Scientific Assessment of Ozone Depletion: 2018; WMO Global Ozone Research and Monitoring Pro-ject—Report No. 58; WMO: Geneva, Switzerland, 2018.
27. IPCC, AR5 Climate Change 2013: The Physical Science Basis; Cambridge University Press: Cambridge, UK, 2013.
28. R. Philipona, C. Mears, M. Fujiwara, P. Jeannet, P. Thorne, G. Bodeker, L. Haimberger, M. Hervo, C. Popp, G. Roma-nens, W. Steinbrecht, R. Stübi, R. Van Malderen, Radiosondes Show That After Decades of Cooling, the Lower Stratosphere Is Now Warming, J. Geophys. Res.: Atmos., 123 (2018) 12,509-512,522.
29. E.R. Kursinski, G.A. Hajj, J.T. Schofield, R.P. Linfield, K.R. Hardy, Observing Earth's atmosphere with radio occultation measurements using the Global Positioning System, J. Geophys. Res.: Atmos., 102 (1997) 23429-23465.
30. A.K. Steiner, F. Ladstädter, C.O. Ao, H. Gleisner, S.P. Ho, D. Hunt, T. Schmidt, U. Foelsche, G. Kirchengast, Y.H. Kuo, K.B. Lauritsen, A.J. Mannucci, J.K. Nielsen, W. Schreiner, M. Schwärz, S. Sokolovskiy, S. Syndergaard, J. Wickert, Consistency and structural uncertainty of multi-mission GPS radio occultation records, Atmos. Meas. Tech., 13 (2020) 2547-2575.
31. H. Gleisner, K.B. Lauritsen, J.K. Nielsen, S. Syndergaard, Evaluation of the 15-year ROM SAF monthly mean GPS radio occultation climate data record, Atmos. Meas. Tech., 13 (2020) 3081-3098.
32. K.Y. Kondratiev, H.I. Niilisk, On the question of carbon dioxide heat radiation in the atmosphere, Geofisica pura e applicata, 46 (1960) 216-230.
33. R.E. Newell, T.G. Dopplick, Questions Concerning the Possible Influence of Anthropogenic $CO_2$ on Atmospheric Temperature, J. Appl. Meteorol. Climatol., 18 (1979) 822-825.
34. S.B. Idso, The Climatological Significance of a Doubling of Earth's Atmospheric Carbon Dioxide Concentration, Science, 207 (1980) 1462-1463.
35. R.S. Lindzen, Can increasing carbon dioxide cause climate change? Proc. Nat. Acad. Sci. (PNAS), 94 (1997) 8335-8342.
36. D.H. Douglass, J.R. Christy, B.D. Pearson, S.F. Singer, A comparison of tropical temperature trends with model predictions, Int. J. Climatol., 28 (2008) 1693-1701.
37. H. Gleisner, M.A. Ringer, S.B. Healy, Monitoring global climate change using GNSS radio occultation, NPJ Clim. Atmos. Sci., 5 (2022) 6.
38. F. Ladstädter, A.K. Steiner, H. Gleisner, Resolving the 21st century temperature trends of the upper troposphere–lower stratosphere with satellite observations, Sci. Rep., 13 (2023) 1306.
39. B. Huang, X. Yin, M. J. Menne, R. Vose, and H. Zhang, NOAA Global Surface Temperature Dataset (NOAAGlobalTemp), Version 6.0.0. NOAA National Centers for Environmental Information.
40. H.E. Brindley, R.P. Allan, Simulations of the effects of interannual and decadal variability on the clear-sky outgoing long-wave radiation spectrum, Q. J. R. Meteorol. Soc., 129 (2003) 2971-2988.

41. J.G. Anderson, J.A. Dykema, R.M. Goody, H. Hu, D.B. Kirk-Davidoff, Absolute, spectrally-resolved, thermal radiance: a benchmark for climate monitoring from space, J. Quant. Spectrosc. Radiat. Transf., 85 (2004) 367-383.
42. K.P. Shine, J.J. Barnett, W.J. Randel, Temperature trends derived from Stratospheric Sounding Unit radiances: The effect of increasing $CO_2$ on the weighting function, Geophys. Res. Lett., 35 (2008).
43. J. Nash, R. Saunders, A review of Stratospheric Sounding Unit radiance observations for climate trends and reanalyses, Q. J. R. Meteorol. Soc., 141 (2015) 2103-2113.
44. L. Wang, C.-Z. Zou, H. Qian, Construction of Stratospheric Temperature Data Records from Stratospheric Sounding Units, J. Clim., 25 (2012) 2931-2946.
45. D.W.J. Thompson, D.J. Seidel, W.J. Randel, C.-Z. Zou, A.H. Butler, C. Mears, A. Osso, C. Long, R. Lin, The mystery of recent stratospheric temperature trends, Nature, 491 (2012) 692-697.
46. C.-Z. Zou, H. Qian, Stratospheric Temperature Climate Data Record from Merged SSU and AMSU-A Observations, J. Atmos. Ocean. Tech., 33 (2016) 1967-1984.
47. D.J. Seidel, J. Li, C. Mears, I. Moradi, J. Nash, W.J. Randel, R. Saunders, D.W.J. Thompson, C.-Z. Zou, Stratospheric temperature changes during the satellite era, J. Geophys. Res.: Atmos., 121 (2016) 664-681.
48. W.J. Randel, A.K. Smith, F. Wu, C.-Z. Zou, H. Qian, Stratospheric Temperature Trends over 1979–2015 Derived from Combined SSU, MLS, and SABER Satellite Observations, J. Clim., 29 (2016) 4843-4859.
49. N. G. Loeb, S.-H. Ham, R. P. Allan, T. J. Thorsen, B. Meyssignac, S. Kato, G. C. Johnson and J. M. Lyman, Surv. Geophys., 45 (2024) 1757-1783.
50. G. Tselioudis, W. B. Rossow, F. Bender, L. Oreopoulos and J. Remillard, Oceanic cloud trends during the satellite era and their radiative signatures, Clim. Dyn., 62 (2024) 9319-9332.